# Anxious Depression Prediction in Real-time Social Data

*Akshi Kumar[a], Aditi Sharma[b], Anshika Arora[c]*

[a,b,c] Dept. of Computer Science & Engineering, Delhi Technological University, Delhi, India-110042

*akshikumar@dce.ac.in

A B S T R A C T

Mental well-being and social media have been closely related domains of study. In this research a novel model, AD prediction model, for anxious depression prediction in real-time tweets is proposed. This mixed anxiety-depressive disorder is a predominantly associated with erratic thought process, restlessness and sleeplessness. Based on the linguistic cues and user posting patterns, the feature set is defined using a 5-tuple vector <*word, timing, frequency, sentiment, contrast*>. An anxiety-related lexicon is built to detect the presence of anxiety indicators. Time and frequency of tweet is analyzed for irregularities and opinion polarity analytics is done to find inconsistencies in posting behaviour. The model is trained using three classifiers (multinomial naïve bayes, gradient boosting, and random forest) and majority voting using an ensemble voting classifier is done. Preliminary results are evaluated for tweets of sampled 100 users and the proposed model achieves a classification accuracy of 85.09%. .

## 1. Introduction

Behavioral psychopathology relates anxiety and depression closely and anxious depression is defined as a mental state of individuals who are diagnosed with depression present in a manner that is more consistent with feeling anxious instead of sad. It is a major depressive disorder (MDD) with a co-morbid anxiety disorder (Wongkoblap, et al. 2017; Almas, et al., 2015; Gaspersz, et al., 2018). Symptoms of both anxiety and depression are of equal intensity and none of them clearly predominates, characterizing a mixed anxiety-depressive disorder. Pertinent studies have linked anxious depression to greater depression severity, reduced treatment response, elevated risk for suicide, and higher risk for cardiovascular disease (Sonawalla, & Fava, 2001).  These findings highlight the importance of expounding issues pertaining to anxious depression as a distinct psychological disorder.

Social media is omnipresent and allows people to self-express, stay connected and in touch with friends and acquaintances across the globe. Social media and mental health of users can be related in three different ways as follows:

- *Social media anxiety disorder:* An active social media user with an addictive pattern, who is distressed by negative interactions and social comparisons on social networking sites affecting self-esteem and mental wellness.
- *Anxious Depression social media verbalization:* An active social media user who uses social media postings as an outlet to share feelings in a non-threatening atmosphere
- *Social Anxiety:* A passive social media user with no engagement and is comfortable with virtual connections rather than real interaction.

Though feelings are hard to articulate but online self-expression provides a means to convey a mental condition into a physical form. Social Media can facilitate pre-diagnosis of a clinical mental health condition related to anxiety, depression or anxious depression in active extroverts who verbalize and share their internal restlessness. The following fig.1 depicts a sample anxious depressive twitter post.

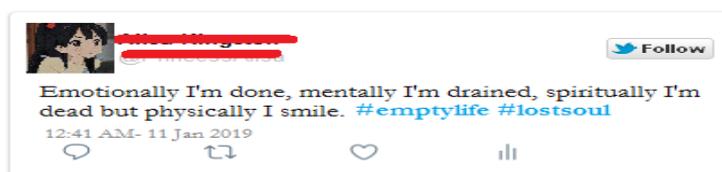

**Fig. 1- Sample Anxious Depression tweet\**

Motivated by this, the research presented in this paper, proffers a predictive model to detect this anxious depression disorder in online twitter posts of 100 sampled users. As the risk of anxiety and depression is higher in youngsters especially teenagers or students who are away from home, we considered the first 100 followers of MS India student forum[*]. The time frame for the posts is taken to be 30 days and linguistic markers along with semantics of posts and users' posting pattern are used to characterize the presence of mixed anxiety depression. The proposed AD (Anxious depression) prediction model is trained using hand-crafted features where the feature vector is a 5-tuple vector <w, t, f, s, c> with each entry representing the following:

- <w:word>: Presence or absence of anxiety related word
- <t:timing>: More than 2 posts during odd hours of night, specifically between 12am to 6am
- <f: frequency>: More than 3 posts in an hour
- <s:sentiment>: More than 25% average posts in 30 days with negative polarity
- <c:contrast>: Presence of more than 25% polarity contrast in posts within the past 24 hours

An anxiety lexicon base has been built as a part of this research which consists of 60 English words suggestive to anxiety. This seed list is grown using the WordNet (Fellbaum, 2002). Tweets are analyzed using this lexicon base for the presence or absence of anxiety related word/words. Further, semantics of the tweet are determined using opinion polarity analytics by SentiWordNet (Baccianella et al., 2010). The users' posting patterns such as odd timing of posts and increased frequency of posts are also investigated to build the model for predicting anxious depression in real-time social data. An Ensemble Vote Classifier, a meta-classifier is used to combine the results of classifiers, namely the Multinomial Naïve Bayes (Kibriya et al., 2004), Gradient Boosting (Friedman, 2002) and Random Forest (Breiman, 2001). The Ensemble Vote Classifier predicts the final class label via majority voting which is the class label that has been predicted most frequently by the classification models. That is, a voting classifier combines the model predictions into ensemble predictions and averages the predictions of the sub-models. The prediction accuracy and F-score of the classifier is used to illustrate the performance of the proposed model. The paper is organized as follows: Section 2 presents the relevant work within the research area depression/anxiety detection in social media posts. The next section, section 3 describes the architecture of proposed AD prediction model followed by accounting of performance results in section 4. Section 5 gives the conclusion of the research with future potential directions.

## 2. Related Work

Various studies have been reported exploring the potential of social media to predict several forms of depression in social media users. The studies are being conducted on the data collected from Twitter (Choudhury et al., 2013a; Choudhury et al., 2013b; Rensik et al., 2015; Pedersen et al., 2015; Tsugawa et al., 2015; Reece et al., 2017; Coppersmith et al., 2014a; Coppersmith et al., 2014b; Nadeem et al., 2016; Wang et al., 2013; Kale, 2015; Park, 2013; Mowery et al., 2017), Facebook (Choudhury et al., 2013c; Schwartz et al., 2014), Reddit (Shen &Rudzicz, 2017 ), Instagram (Reece & Danforth, 2017) and web forums.

In a study (Shen &Rudzicz, 2017) authors classified Reddit posts related to anxiety by applying apply N-gram language modeling, vector embeddings, topic analysis, and emotional norms to generate features. In a work (Choudhury et al., 2013a, Choudhury et al., 2013b) authors used crowd sourcing to collect data of Twitter users with clinical depression and measure behavioural attributes to build a classifier in order to identify depression in a person. In another study (Choudhury et al., 2013c) the authors used Facebook data and developed a series of statistical models to predict postpartum depression. Authors (Rensik et al., 2015) explored the use of supervised topic models in the analysis of linguistic signal for detecting depression in Twitter. Authors (Pedersen et al. 2015) used lexical features to train decision lists to identify Twitter users with depression. In another work (Tsugawa et al., 2015) authors extracted features from activity history of users on Twitter for estimating the degree of depression. In a research (Schwartz et al. 2014) authors used survey responses and status updates from 28,749 Facebook users to develop a regression model that predicts users' degree of depression based on their Facebook status updates. Authors (Reece et al., 2017) extracted predictive features measuring affect, linguistic style, and context from Twitter data to build models with supervised learning algorithms in order to predict emergence of depression and Post-Traumatic Stress Disorder in Twitter users. In a study (Coppersmith et al., 2014a) authors presented a method to obtain a post-traumatic stress disorder classifier for social media using Twitter data demonstrating its utility by examining differences in language use between PTSD and random individuals. In another study (Coppersmith et al., 2014b) the authors analyzed mental health phenomena in publicly available Twitter data by gathering data for a range of mental illnesses quickly and cheaply to detect post-traumatic stress disorder, depression, bipolar disorder, and seasonal affective disorder. Authors (Nadeem et al., 2016) used crowd sourcing to collect data of Twitter users with diagnosed depression. They used Bag of words approach to quantify each tweet in order to leverage several statistical classifiers to provide estimates to the risk of depression. (Wang et al., 2013) proposed a depression detection model based on 10 features and 3 classifiers to verify the model on Twitter data. They also developed an application based on their proposed model for mental health monitoring online. In another work (Reece & Danforth, 2017) authors applied machine learning tools to identify markers of depression on photos from Instagram. They used color analysis, metadata components, and algorithmic face detection to extract statistical features from Instagram photos. Author (Kale, 2015) targeted Twitter data to identify users who could potentially suffer from mental disorders, and classify them based on the intensity of linguistic usage and different behavioural features using sentiment analysis techniques. In a work (Park, 2013) authors concluded in their work that Non-depressed individuals perceived Twitter as an information consuming and sharing tool, while depressed individuals perceived it as a tool for social awareness and emotional interaction. In a work (Mowery et al., 2017) authors conducted two feature study experiments in order to use the predictive power of supervised machine learning classifiers and study the influence of feature sets for classifying depression-related tweets.

Recently deep learning techniques have also been used for depression detection on social media. Authors (Orabi et al., 2018) used deep neural networks on Twitter data for detection of depression among Twitter users. (Trotzek et al., 2018) compared their convolution neural network-based model on different word embeddings to a classification based on user-level linguistic metadata. Finally, an ensemble of both approaches is shown to achieve state-of-the-art results in a depression detection task.

---

[*]https://twitter.com/msindiastudent

Various secondary studies have also given insights to the models built for identifying depression on social media like Facebook, Twitter, and web forums (Guntuku et al., 2017; Seabrook, 2016; Wongkoblap, 2017).

## 3. The Proposed Anxious Depression (AD) Prediction Model

Anxious depression is a mental health concern where pre-diagnosis, which is early screening of symptoms, may help signal warning about the degree of disorder. The pervasive social web can provide an apposite test-bed for understanding user behaviour and mental health. This research proffers a predictive model to detect anxious depression in real-time tweets of users. The following figure 2 depicts the architecture of the proposed AD prediction model.

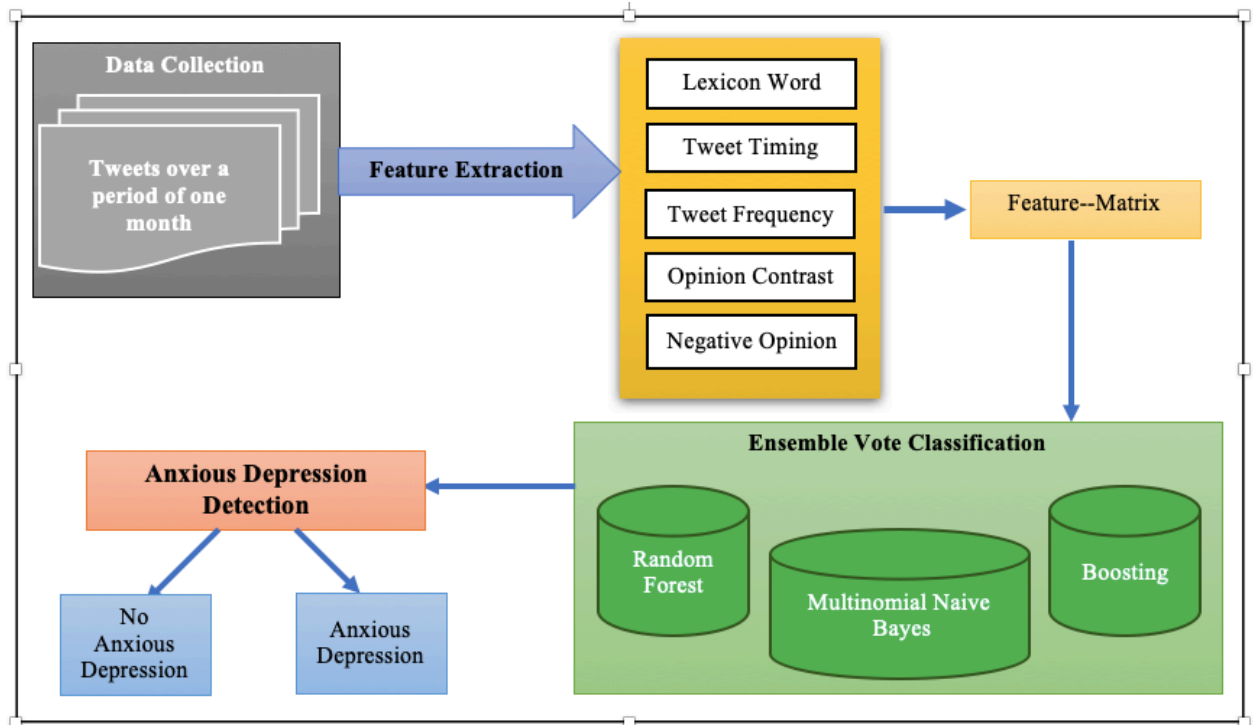

**Fig.2- System Architecture for AD Prediction Model**

The following sub-sections explicate the details of the model:

*3.1 Data Collection*
The dataset with past one month tweets of 100 sampled users is scrapped using the Twitter API. The first 100 followers of MS India student forum are considered for this research. Each user's data consists of name, date of account creation, account verification status (verified or not), language, description and tweet count. For each user, tweets are fetched, with date and time of post, number of re-tweets, hash tags, mentioned users.

*3.2 Pre-processing*
Pre-processing is the process of cleaning and filtering the data to make it suitable for the feature extraction. The process includes:
- Removing numeric and empty texts, URLs, mentions, hashtags, non-ASCII characters, stop-words[†] and punctuations
- Tokenization of tweets is done using the TreebankWordTokenizer of Python Natural Language Toolkit (NLTK)[‡] to filter the words, symbols and other elements called tokens(Loper,& Bird, 2002) The tokens are converted to lower case.

---

[†] University of Glasgow Stop-word list
[‡] https://www.nltk.org/

- Replacing slangs and emojis by their descriptive text using the *SMS Dictionary*[§] and *emojipedia*[**] respectively. As Internet is an informal way of communication, the use of slangs and emojis is a common practice. These may help understand the context and also intensify the emotion associated.
- Stemming to reduce the words to their root words using Porter's stemmer[6]. Stemming enhances the likelihood of matching to the lexicon. Tweets can at most contain 280 characters, so users tend to write in short forms. Different users can use different terms for the same word, not every synonym word can be included to lexicon, it will increase the processing time, stemming is crucial for the accuracy of the prediction model.

*3.3 Feature Engineering*

The feature vector for building the learning model is trained using a 5-tuple vector <w, t, f, s, c>, where,

- <w:word>: Presence or absence of anxiety related word using the anxiety lexicon base
- <t:timing>: More than 2 posts during odd hours of night, specifically between 12am to 6am
- <f:frequency>: More than 3 posts in an hour during anytime of the day
- <s:sentiment>: More than 25% average posts in 30 days with negative polarity
- <c:contrast>: Presence of more than 25% polarity contrast in posts within the past 24 hours

The details of feature extraction and method of feature value assignment to generate feature vector are as follows:

*3.3.1 Anxiety Lexicon Base*

Depressive rumination is the compulsive focus of attention on thoughts that cause feelings of sadness, anxiety and distress. A person with anxious depression disorder is very likely to verbalize thoughts using specific anxiety related words. Therefore, an anxiety lexicon base with a seed list of 60 words is built with keywords that represent anxious depression in textual content. The seed list is eventually grown using WordNet. The processed tokens from the tweets are matched with the lexicon and the feature value is set to true ('1') if the word is present in the lexicon base else it is set to '0'. Table 1 presents the lexicon base of initial 60 words.

**Table 1-Lexicon for Anxiety Detection**

| Anxious depression related words |
|---|
| Fat, bad, weak, problem, tired, illusion, restless, bored, crap, shit, fuck, sad, escape, useless, meaningless, crying, reject, suffer, sleepless, never, bored, afraid, unhappy, ugly, upset, awful, torture, unsuccessful, helpless, suffer, fail, sorrow, nobody, blame, damaged, shatter. pathetic, insomnia, kill, panic, lonely, hate, depressed, frustrated, loser, suicidal, hurt, painful, disappoint, broke, abandon, worthless, regret, dissatisfied, lost, empty, destroy, ruin, die, sick. |

*3.3.2 Tweet Timing*

Chronic insomnia, which is sleeplessness, is one of the most common symptoms of anxious depression and stress. Users who are active through the midnight hours, i.e. from 12 am to 6 am, evidently show psychological disturbance in sleep pattern with increased restlessness and over-thinking. Therefore, tweet timing is an important feature and the value of feature is set to 1 if two or more than 2 tweets are posted after mid-night during odd hours of 12am to 6am, else it is set to 0.

*3.3.3 Tweet Frequency*

Though timing of tweets is primarily associated with odd-hour postings, generic tweet frequency within 24 hours also demonstrates user's restlessness and urge to share. The feature is set to true: '1' if the no. of tweets in any hour of the day is equal to or greater than 3, else it is set to false: '0'.

*3.3.4 Negative Polarity tweets*

Opinion polarity takes into account the amount of positive or negative terms that appear in a given text (Kumar & Sebastian, 2012a; Kumar & Sebastian, 2012b; Kumar & Sebastian; 2012c, Kumar et al. 2015; Kumar & Joshi, 2017a; Kumar & Joshi, 2017b; Kumar & Jaiswal, 2017; Kumar & Khorwal, 2017; Kumar & Sharma, 2017; Son et al., 2019; Kumar & Jaiswal, 2019; Kumar & Garg, 2019). In this research, to determine the polarity of tweets, SentiWordNet is used. SentiWordNet is a lexical resource which assigns the polarity to the words. The words of the WordNet are classified into the synset, and then each synset is assigned three values between the range of 0.0 and 1.0 representing the positive, negative and neutral polarity of the word. A single word can depict different sentiments in different scenarios, so as the word can have all three polarities of non-zero value. A cumulative value signifying the average of negative polarity tweets posted within the considered 30 days time frame is calculated. The feature is set to a value of '1' if 25% or more tweets posted have negative polarity; else it is set to '0'.

*3.3.5 Polarity Contrast*

The shift or contrast in polarity of posts from negative to positive or positive to negative is indicative of inconsistent mental state and restlessness. Typically described as a flip-flop behaviour, the person with anxious depression disorder often changes opinions and has confused thinking. Thus, to calculate this contrast *c* between opinion polarities of tweets, the following equation (1) is used:

---

[§]SMS Dictionary. *Vodacom Messaging*. Retrieved 16 March 2012.
[**]https://emojipedia.org/
[6]Porter Stemmer: http://snowball.tartarus.org/

*http://ssrn.com/link/ICAESMT-2019.html=xyz*
*Information Systems &eBusiness Network (ISN)*

$$c = \frac{(\delta \cdot PP + pw) - (\delta \cdot NP + nw)}{(\delta \cdot PP + pw) + (\delta \cdot NP + nw)} \quad (1)$$

where,
pw is the count of words with positive opinion polarity
nw is the count of words with negative opinion polarity
PP is the count of positive post
NP is the count of negative post
δ is the post co-efficient, the value of which is set to 3.

If a polarity contrast of >=0.25 magnitude is observed in tweets, then the feature value is set to '1'; else to '0'.

*3.4. Supervised Learning*

The model is now trained using these features. Three machine learning (Bhatia & Kumar, 2008; Kumar & Jaiswal, 2017; Kumar & Sangwan, 2019) classifiers are used namely, Multinomial Naïve Bayes, Gradient Boosting and Random Forest. An ensemble vote classifier with majority voting mechanism is used to generate the final prediction. The data split for training and testing was 80 and 20.10 fold cross-validation was done.

## 4. Results

Accuracy and F-score measures are used to evaluate the performance of the ensemble vote classifier. The accuracy of the ensemble vote predictive model is compared to the three individual classifiers. The following Table 2 gives the accuracy of the individual classifiers and the ensemble vote classifier.

**Table 2-**Classification accuracy of classifiers

| Classifier | Accuracy (%) |
|---|---|
| Multinomial Naïve Bayes | 77.89 |
| Random Forest | 81.04 |
| Gradient Boosting | 79.12 |
| Ensemble Vote Classifier | 85.09 |

The following Figure 3 illustrates the comparative performances of the classifiers graphically

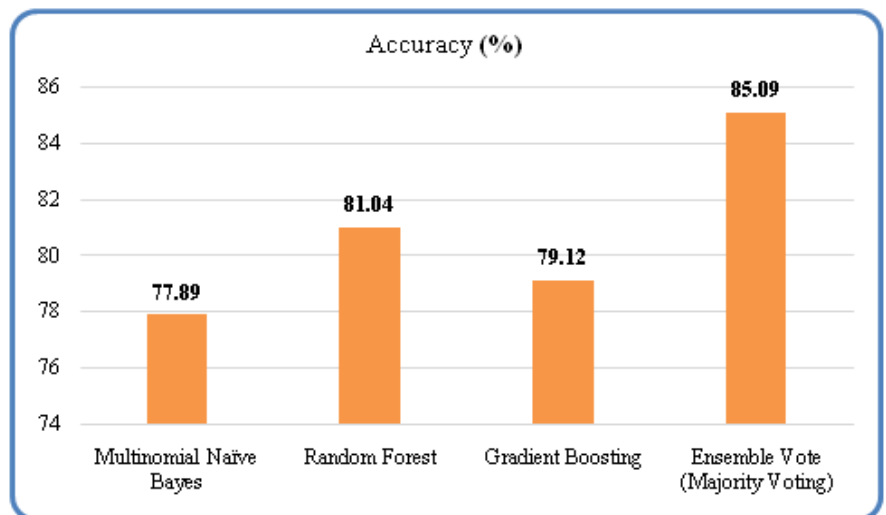

**Fig.3. Accuracy of classifiers**

The accuracy of the proposed AD prediction model is 85.09% with an F-score of 79.68%. The model is able to achieve motivating results and predicts users with anxious depression disorder. The following figure 4depicts the performance results graphically.

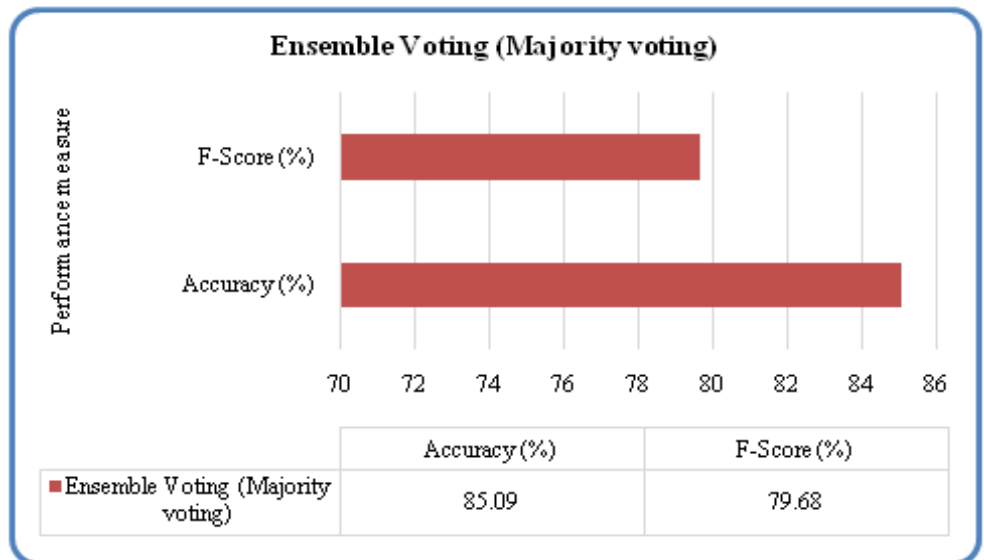

**Fig.4. Performance of proposed AD prediction model**

## 5. Conclusion and Future work

Social media has revolutionized the way we interact with the world, allowing us all to stay connected and self-express. Mixed anxiety depression and social media seem to exist in a vicious cycle with one problem often stimulates the other. A supervised learning based prediction model is proposed in this research, where tweets of first 100 followers of MS India student forum are analyzed using various linguistic, semantic and activity features to detect anxious depression disorder. The presence of anxiety related words were considered as linguistic markers whereas count of negative tweets and polarity contrast of tweets d\related to semantic marker. Users' post timing and frequency were also considered to build a model to efficiently predict anxious depression in users. Nearly 85% predictions are found to be accurate in the preliminary analysis. As a possible future work, fine grain emotion analysis can be done to detect anxiety indicators instead of using SentiWordNet which categorizes the words into three polarities. Further, the model can be tested on different user base: geographic, age, profession etc.  Neuro-fuzzy and deep learning models can be explored for superlative prediction performance.